\documentclass[aps,prl,twocolumn,superscriptaddress,showpacs,floatfix]{revtex4} 

\usepackage{graphicx}
\usepackage{dcolumn}
\usepackage{bm}
\usepackage{url} 
\usepackage{amsmath} 
\usepackage{amssymb} 

\begin{document}

\title{Search for High-Mass \boldmath$e^+e^-$ Resonances in \boldmath$p\bar{p}$ Collisions at \boldmath$\sqrt{s}=$1.96 TeV}

\affiliation{Institute of Physics, Academia Sinica, Taipei, Taiwan 11529, Republic of China} 
\affiliation{Argonne National Laboratory, Argonne, Illinois 60439} 
\affiliation{University of Athens, 157 71 Athens, Greece} 
\affiliation{Institut de Fisica d'Altes Energies, Universitat Autonoma de Barcelona, E-08193, Bellaterra (Barcelona), Spain} 
\affiliation{Baylor University, Waco, Texas  76798} 
\affiliation{Istituto Nazionale di Fisica Nucleare Bologna, $^w$University of Bologna, I-40127 Bologna, Italy} 
\affiliation{Brandeis University, Waltham, Massachusetts 02254} 
\affiliation{University of California, Davis, Davis, California  95616} 
\affiliation{University of California, Los Angeles, Los Angeles, California  90024} 
\affiliation{University of California, San Diego, La Jolla, California  92093} 
\affiliation{University of California, Santa Barbara, Santa Barbara, California 93106} 
\affiliation{Instituto de Fisica de Cantabria, CSIC-University of Cantabria, 39005 Santander, Spain} 
\affiliation{Carnegie Mellon University, Pittsburgh, PA  15213} 
\affiliation{Enrico Fermi Institute, University of Chicago, Chicago, Illinois 60637} 
\affiliation{Comenius University, 842 48 Bratislava, Slovakia; Institute of Experimental Physics, 040 01 Kosice, Slovakia} 
\affiliation{Joint Institute for Nuclear Research, RU-141980 Dubna, Russia} 
\affiliation{Duke University, Durham, North Carolina  27708} 
\affiliation{Fermi National Accelerator Laboratory, Batavia, Illinois 60510} 
\affiliation{University of Florida, Gainesville, Florida  32611} 
\affiliation{Laboratori Nazionali di Frascati, Istituto Nazionale di Fisica Nucleare, I-00044 Frascati, Italy} 
\affiliation{University of Geneva, CH-1211 Geneva 4, Switzerland} 
\affiliation{Glasgow University, Glasgow G12 8QQ, United Kingdom} 
\affiliation{Harvard University, Cambridge, Massachusetts 02138} 
\affiliation{Division of High Energy Physics, Department of Physics, University of Helsinki and Helsinki Institute of Physics, FIN-00014, Helsinki, Finland} 
\affiliation{University of Illinois, Urbana, Illinois 61801} 
\affiliation{The Johns Hopkins University, Baltimore, Maryland 21218} 
\affiliation{Institut f\"{u}r Experimentelle Kernphysik, Universit\"{a}t Karlsruhe, 76128 Karlsruhe, Germany} 
\affiliation{Center for High Energy Physics: Kyungpook National University, Daegu 702-701, Korea; Seoul National University, Seoul 151-742, Korea; Sungkyunkwan University, Suwon 440-746, Korea; Korea Institute of Science and Technology Information, Daejeon, 305-806, Korea; Chonnam National University, Gwangju, 500-757, Korea} 
\affiliation{Ernest Orlando Lawrence Berkeley National Laboratory, Berkeley, California 94720} 
\affiliation{University of Liverpool, Liverpool L69 7ZE, United Kingdom} 
\affiliation{University College London, London WC1E 6BT, United Kingdom} 
\affiliation{Centro de Investigaciones Energeticas Medioambientales y Tecnologicas, E-28040 Madrid, Spain} 
\affiliation{Massachusetts Institute of Technology, Cambridge, Massachusetts  02139} 
\affiliation{Institute of Particle Physics: McGill University, Montr\'{e}al, Canada H3A~2T8; and University of Toronto, Toronto, Canada M5S~1A7} 
\affiliation{University of Michigan, Ann Arbor, Michigan 48109} 
\affiliation{Michigan State University, East Lansing, Michigan  48824}
\affiliation{Institution for Theoretical and Experimental Physics, ITEP, Moscow 117259, Russia} 
\affiliation{University of New Mexico, Albuquerque, New Mexico 87131} 
\affiliation{Northwestern University, Evanston, Illinois  60208} 
\affiliation{The Ohio State University, Columbus, Ohio  43210} 
\affiliation{Okayama University, Okayama 700-8530, Japan} 
\affiliation{Osaka City University, Osaka 588, Japan} 
\affiliation{University of Oxford, Oxford OX1 3RH, United Kingdom} 
\affiliation{Istituto Nazionale di Fisica Nucleare, Sezione di Padova-Trento, $^x$University of Padova, I-35131 Padova, Italy} 
\affiliation{LPNHE, Universite Pierre et Marie Curie/IN2P3-CNRS, UMR7585, Paris, F-75252 France} 
\affiliation{University of Pennsylvania, Philadelphia, Pennsylvania 19104}
\affiliation{Istituto Nazionale di Fisica Nucleare Pisa, $^y$University of Pisa, $^z$University of Siena and $^{aa}$Scuola Normale Superiore, I-56127 Pisa, Italy} 
\affiliation{University of Pittsburgh, Pittsburgh, Pennsylvania 15260} 
\affiliation{Purdue University, West Lafayette, Indiana 47907} 
\affiliation{University of Rochester, Rochester, New York 14627} 
\affiliation{The Rockefeller University, New York, New York 10021} 
\affiliation{Istituto Nazionale di Fisica Nucleare, Sezione di Roma 1, $^{bb}$Sapienza Universit\`{a} di Roma, I-00185 Roma, Italy} 
\affiliation{Rutgers University, Piscataway, New Jersey 08855} 
\affiliation{Texas A\&M University, College Station, Texas 77843} 
\affiliation{Istituto Nazionale di Fisica Nucleare Trieste/Udine, $^{cc}$University of Trieste/Udine, Italy} 
\affiliation{University of Tsukuba, Tsukuba, Ibaraki 305, Japan} 
\affiliation{Tufts University, Medford, Massachusetts 02155} 
\affiliation{Waseda University, Tokyo 169, Japan} 
\affiliation{Wayne State University, Detroit, Michigan  48201} 
\affiliation{University of Wisconsin, Madison, Wisconsin 53706} 
\affiliation{Yale University, New Haven, Connecticut 06520} 
\author{T.~Aaltonen}
\affiliation{Division of High Energy Physics, Department of Physics, University of Helsinki and Helsinki Institute of Physics, FIN-00014, Helsinki, Finland}
\author{J.~Adelman}
\affiliation{Enrico Fermi Institute, University of Chicago, Chicago, Illinois 60637}
\author{T.~Akimoto}
\affiliation{University of Tsukuba, Tsukuba, Ibaraki 305, Japan}
\author{M.G.~Albrow}
\affiliation{Fermi National Accelerator Laboratory, Batavia, Illinois 60510}
\author{B.~\'{A}lvarez~Gonz\'{a}lez}
\affiliation{Instituto de Fisica de Cantabria, CSIC-University of Cantabria, 39005 Santander, Spain}
\author{S.~Amerio$^x$}
\affiliation{Istituto Nazionale di Fisica Nucleare, Sezione di Padova-Trento, $^x$University of Padova, I-35131 Padova, Italy} 
\author{D.~Amidei}
\affiliation{University of Michigan, Ann Arbor, Michigan 48109}
\author{A.~Anastassov}
\affiliation{Northwestern University, Evanston, Illinois  60208}
\author{A.~Annovi}
\affiliation{Laboratori Nazionali di Frascati, Istituto Nazionale di Fisica Nucleare, I-00044 Frascati, Italy}
\author{J.~Antos}
\affiliation{Comenius University, 842 48 Bratislava, Slovakia; Institute of Experimental Physics, 040 01 Kosice, Slovakia}
% ...

\collaboration{CDF Collaboration\footnote{With visitors from 
$^a$University of Massachusetts Amherst, Amherst, Massachusetts 01003,
$^b$Universiteit Antwerpen, B-2610 Antwerp, Belgium, 
$^c$University of Bristol, Bristol BS8 1TL, United Kingdom,
$^d$Chinese Academy of Sciences, Beijing 100864, China, 
$^e$Istituto Nazionale di Fisica Nucleare, Sezione di Cagliari, 09042 Monserrato (Cagliari), Italy,
$^f$University of California Irvine, Irvine, CA  92697, 
$^g$University of California Santa Cruz, Santa Cruz, CA  95064, 
$^h$Cornell University, Ithaca, NY  14853, 
$^i$University of Cyprus, Nicosia CY-1678, Cyprus, 
$^j$University College Dublin, Dublin 4, Ireland,
$^k$Royal Society of Edinburgh/Scottish Executive Support Research Fellow,
$^l$University of Edinburgh, Edinburgh EH9 3JZ, United Kingdom, 
$^m$Universidad Iberoamericana, Mexico D.F., Mexico,
$^n$Queen Mary, University of London, London, E1 4NS, England,
$^o$University of Manchester, Manchester M13 9PL, England, 
$^p$Nagasaki Institute of Applied Science, Nagasaki, Japan, 
$^q$University of Notre Dame, Notre Dame, IN 46556,
$^r$University de Oviedo, E-33007 Oviedo, Spain, 
$^s$Simon Fraser University, Vancouver, British Columbia, Canada V6B 5K3,
$^t$Texas Tech University, Lubbock, TX  79409, 
$^u$IFIC(CSIC-Universitat de Valencia), 46071 Valencia, Spain,
$^v$University of Virginia, Charlottesville, VA  22904,
$^{dd}$On leave from J.~Stefan Institute, Ljubljana, Slovenia, 
}}
\noaffiliation
\begin{abstract}
A search for high-mass resonances in the $e^+e^-$ final state is presented
based on 2.5 fb$^{-1}$ of $\sqrt{s}=$1.96 TeV $p\bar{p}$ collision data from
the CDF II detector at the Fermilab Tevatron. The largest excess over the
standard model prediction is at an $e^+e^-$ invariant mass of 240
GeV/$c^2$. The probability of observing such an excess arising from
fluctuations in the standard model anywhere in the mass range of 150--1,000
GeV/$c^2$ is 0.6\% (equivalent to 2.5 $\sigma$). We exclude the standard model
coupling $Z'$ and the Randall-Sundrum graviton for $k/\overline{M}_{Pl}=0.1$
with masses below 963 and 848 GeV/$c^2$ at the 95\% credibility level,
respectively.
\end{abstract}
\maketitle
%%%%%%%%%%%%%%%%%%
The charged lepton-antilepton pair signature, in particular $e^+e^-$ and
$\mu^+\mu^-$, has been a leading discovery channel for new particles such as
the $J/\psi$ and $\Upsilon$ mesons and the $Z$ boson since they have cleaner
experimental signatures and lower backgrounds than hadronic signatures.

Many models beyond the standard model (SM) predict the existence of new
particles decaying to lepton-antilepton pairs. The $E_6$
$Z^{\prime}$s~\cite{zprime} and the Randall-Sundrum (RS) graviton~\cite{RSG}
are examples of specific new particles decaying to a lepton-antilepton final
state. The $Z'_{\psi}$, $Z'_{\chi}$, $Z'_{\eta}$, $Z'_{I}$, $Z'_{sec}$, and
$Z'_{N}$ are chosen for testing the $E_6$ model. Assuming one extra dimension,
we test the RS model in the $k/\overline{M}_{Pl}$ range between 0.01 and
0.1~\cite{RSG_RANGE}, where $k$ is the curvature of the extra dimension and
$\overline{M}_{Pl}$ is the reduced effective Planck scale.

In recent publications, the CDF~\cite{CDFZP} and D0~\cite{D0RS} Collaborations
set limits on these models with 1.3 and 1.0 fb$^{-1}$ of integrated luminosity
and limits on $Z'$ with SM coupling and RS graviton for
$k/\overline{M}_{Pl}$=0.1 are 923 and 900 GeV/$c^2$, respectively. Using a data
set twice as large (2.5 fb$^{-1}$), this Letter describes a search for $e^+e^-$
resonances in the invariant mass range of 150--1,000 GeV/$c^2$, and we set
upper limits on $\sigma(p\bar{p}\rightarrow X)\cdot\mathcal{B}(X\rightarrow
e^+e^-)$ at the 95\% credibility level (C.L.) where $X$ is a spin 1 or spin 2
particle. We also set lower mass bounds on the $Z'$ with SM coupling, the $Z'$s
in the $E_6$ model, and the RS graviton. 

This analysis is based on data collected with the CDF II
detector~\cite{CDF}. The relevant components of the detector for this analysis
are the tracking system and the calorimeters. The tracking system consists of a
96 layer drift chamber called the central outer tracker (COT), surrounding an
eight-layer silicon tracker. Both are inside a 1.4 T solenoidal magnet. The COT
covers the range of pseudorapidity $|\eta|<$1.1~\cite{cdfcoord}, and the
silicon tracker covers $|\eta|$ up to 2.0. The electromagnetic (EM) and
hadronic calorimeters, which are sandwiches of lead (EM) or iron (hadronic)
absorber and plastic scintillator. They are outside the magnet, and are divided
into a central calorimeter ($|\eta|<$1.1) and two plug calorimeters
(1.1$<|\eta|<$3.6). Both the central and the plug EM calorimeters have
fine-grained shower profile detectors at EM shower maximum. 

We use the same on-line event selection criteria (triggers) used in our
previous report~\cite{CDFZP}. Off-line events are required to have two isolated
electrons~\cite{ccong}, one in the central EM calorimeter and the other one in
either the central (CC) or the plug (CP) EM calorimeters. Only electrons with
$E_T$~\cite{cdfcoord} greater than 25 GeV and $|\eta|<$ 2 are used in order to
ensure 100\% trigger efficiency and coverage by the the silicon
tracker. Electrons in the central EM calorimeter are required to have a
well-measured track in the COT system pointing at an energy deposit in the
calorimeter. For electrons in the plug EM calorimeter, the track association
uses a calorimeter-seeded silicon-tracking algorithm~\cite{PHX}. An
opposite-charge requirement is applied to electron-objects pairs detected in
the central EM calorimeter. No such requirement is applied when one electron is
detected in the plug, where $\eta$-dependent charge misidentification
occurs. Events with both electrons in the plug EM calorimeter are not
considered in this Letter since adding them gains little sensitivity. 

The {\sc pythia}~\cite{pythia} Tune A~\cite{tuneA} Monte Carlo event generator
is used to model the expected signals and backgrounds unless otherwise
stated. For spin 1 $Z'$, SM-like couplings are assumed, and for spin 2
resonances, the RS graviton model with $k/\overline{M}_{Pl}$=0.1 is used. The
total selection efficiencies of spin 1 particles vary from 27 to 38\% and those
of spin 2 particles vary from 28 to 32\% as functions of the particle mass in
the search range. 

There are three sources of background. One is Drell-Yan production of $e^+e^-$
pairs (DY), which is the dominant source of background and is
irreducible. Another is dijets and $W$+jets production (referred to as ``QCD''
background) where one or more jets is misidentified as electron. Other
contributions include $Z/\gamma^{*}\rightarrow\tau^+\tau^-$, $t\bar{t}$, and
diboson ($W\gamma,WW,WZ,ZZ,\gamma\gamma$) production that collectively are
referred to as ``other SM'' backgrounds. 

The simulated DY prediction is normalized to the data after subtracting other
SM and QCD backgrounds in an invariant mass window from 76 to 106 GeV/$c^2$ for
CC events and from 81 to 101 GeV/$c^2$ for CP events to estimate the DY
background. Different mass windows are used because the QCD background rate in
CP events is higher than in CC events. We assign a 3.6\% systematic uncertainty
in the DY prediction to take into account the invariant-mass dependence of the
$k$-factor~\cite{kfactor} that is the difference between the leading and the
next-to-next-to-leading order DY cross sections. The uncertainty in the DY
prediction due to the choice of the parton distribution function set
CTEQ6M~\cite{cteq6m} using the Hessian method~\cite{pdf} is 3.7$-$6.4$-$13\%
(200$-$600$-$1,000 GeV/$c^2$) depending on the invariant mass.

The QCD background estimation is determined from the experimental data. The
estimate is obtained using the probability for a jet to be misidentified as an
electron~\cite{rfake}. We measure this probability with a jet-triggered data
sample. We then apply the misidentification probability to each jet in events
with one good electron candidate and one or more jets. To estimate the dijet
background contribution, events with $W$ or $Z$ candidates are removed from the
sample before applying the jet misidentification probability (MP). Events with
$W$ candidates are identified with one good electron and a large missing
transverse energy ${\not\!\!E_T}$~\cite{MET} and events with $Z$ candidates are
identified with two ``loose electrons''. To estimate the $W$+jets background,
events with $Z$ candidates are removed and events with $W$ candidates are
retained. The dominant systematic uncertainty in the predicted QCD background
is due to the 20\% uncertainty in the jet MP, which is obtained from the
variation in the MP measured in the different jet data samples.

\begin{figure}[ht]
\includegraphics[height=2.5in,width=3.5in]{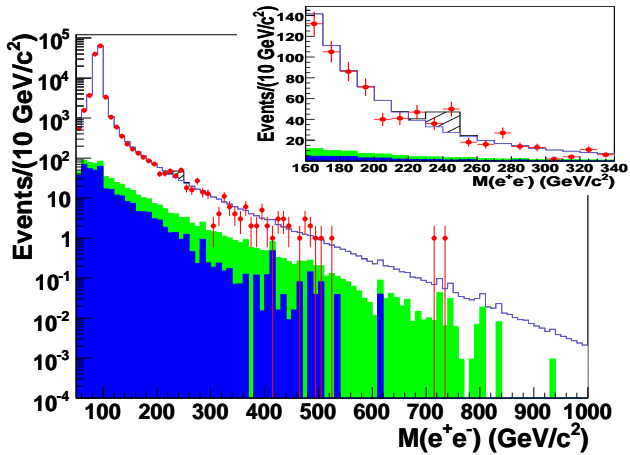}
\caption{Invariant mass distribution of $e^+e^-$ events compared to the
  expected backgrounds. Dots with error bars are data. The dark shaded region
  represents ``other SM'' background, the light shaded region shows ``QCD''
  background, and the white region corresponds to $Z/\gamma^{*}\rightarrow
  e^+e^-$ background. The inset shows the same for the 240 GeV/$c^2$
  region. The hatched histogram shows the shape of the expected signal from a
  240 GeV/$c^2$ spin 1 particle (of negligible intrinsic width) on top of the
  total background. The hatched region is normalized to the number of excess
  events seen in the data.}\label{fig:mee}
\end{figure}

Other SM contributions to the background are estimated with simulation
samples~\cite{WGAMMA}. These simulated samples are normalized to the product of
the theoretical cross sections and the integrated
luminosity. Fig.~\ref{fig:mee} shows the observed $e^+e^-$ invariant mass
spectrum from 2.5 fb$^{-1}$ of data together with the expected backgrounds.
%%%%%%%%%%%%%%%%%%
%%%%%%%%%%%%%%%%%%

The systematic uncertainty for other SM backgrounds is dominated by the 6\%
uncertainty in the integrated luminosity measurement~\cite{Lumi} and 8\%
uncertainty in the theoretical cross sections~\cite{TheoryX}. Other systematic
sources are the uncertainty on the scale factor of electron identification
efficiency that comes from the difference between data and simulated events
(1.3\% for CC and 2.3\% for CP events), the energy scale (1.0\%), and the
energy resolution (0.6\% for CC and 0.3\% for CP events), which affects the
shape of the $e^+e^-$ invariant mass distribution. The uncertainty on the
acceptance due to parton-distribution-function uncertainties is evaluated using
the same method that was used for the DY prediction, and found to be 1.9\% for
CC and 0.6\% for CP events.

The search for $e^+e^-$ resonances in the high-mass range of 150--1,000
GeV/$c^2$ uses an unbinned likelihood ratio statistic, $\lambda$, defined in
Eqs.~\ref{eq:rlike1}$-$\ref{eq:rlike3}~\cite{pvalue}:
\begin{equation}
    \lambda=\frac{\underset{n_b\geq0}{max.}\mathcal{L}_b}{\underset{n_b\geq0,n_s\geq0}{max.}\mathcal{L}_{s+b}},~~0\leq\lambda\leq1,~~0\leq-2\ln\lambda\leq\infty 
  \label{eq:rlike1}
\end{equation}
\begin{equation}
    \mathcal{L}_{s+b}=\frac{(n_s+n_b)^N
    e^{-(n_s+n_b)}}{N!}\prod^N_i\frac{n_s S(x_i|\mu)+n_b B(x_i)}{n_s+n_b}
  \label{eq:rlike2}
\end{equation}
\begin{equation}
  \mathcal{L}_{b}=\frac{n_b^N e^{-n_b}}{N!}\prod^N_i B(x_i).
  \label{eq:rlike3}
\end{equation}
where $\mathcal{L}_{b}$ is the likelihood for a null hypothesis that is
described by the SM only, $\mathcal{L}_{s+b}$ is the likelihood for a test
hypothesis that is described by physics beyond the SM together with the SM. The
quantities $n_s$ and $n_b$ are the number of signal and background candidates
which are determined by the fit and $N$ is the number of candidates observed in
data, each represented by a vector $\{x_i\}$ of observables. The signal
probability density function (PDF), $S(x|\mu)$, is a Gaussian with a floating
mean $\mu$ and a fixed width, and $B(x)$ is a background PDF obtained from the
total background template. The widths of the signal PDF are determined from
simulation ($\sigma_{M_{ee}}=0.8565$ GeV/$c^2$ $+0.0192\cdot M_{ee}$ for
$M_{ee}>150$ GeV/$c^2$) with the assumption that the decay widths of resonances
are much less than the experimental resolution. The quantities
$\mathcal{L}_{s+b}$ and the $\mathcal{L}_{b}$ are maximized separately without
external background constraints. The function $-2\ln\lambda$ is calculated over
the search range of 150--1,000 GeV/$c^2$ and the most prominent local maxima
are listed in Table~\ref{table:localmax}. The most significant deviation
between data and the SM prediction occurs at an invariant mass of 241.3
GeV/$c^2$ where $-2\ln\lambda$ is 14.4. The
$\rm{(data-background)}/\sigma_B$~\cite{SOB} corresponding to the region of
maximum $-2\ln\lambda$ is calculated by counting the number of observed events
and estimated backgrounds within $\pm$2 $\sigma_{M_{ee}}$ of the maximum, and
it is 3.8. 
%%%%%%%%%%%%%%%%%%
\begin{table}[htbp]
  \begin{center}
  \caption{The prominent local maxima in the search range of 150--1,000
  GeV/$c^2$.} \label{table:localmax}
  \begin{tabular}{ccccc} \hline \hline
    $M_{X}$ (GeV/$c^2$)&241.3&272.7&478.9&725.2 \\
    $-2\ln\lambda$    &14.4&3.7 &2.6 &4.1 \\ \hline \hline
  \end{tabular}
  \end{center}
\end{table}
%%%%%%%%%%%%%%%%%%

\begin{figure}[htbp]
\includegraphics[height=2.3in,width=3.5in]{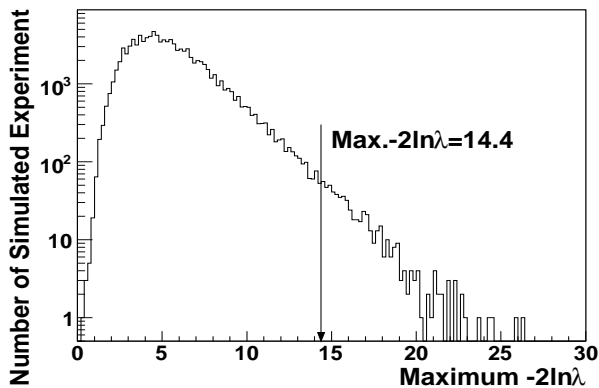}
\caption{Distribution of maximum $-2\ln\lambda$ in simulated experiments that
  assume only background. The arrow indicates the value observed in data:
  $-2\ln\lambda$=14.4.} \label{fig:rlike}
\end{figure}

To estimate the probability of observing an excess equal to or greater than the
maximum observed excess anywhere in the search range of 150--1,000 GeV/$c^2$,
we simulated 100,000 experiments assuming background only. The distribution of
maximum $-2\ln\lambda$ on these simulated experiments is shown in
Fig.~\ref{fig:rlike}. Assuming only SM physics, the probability of observing a
number of events equal to or greater than the observed excess is defined as the
fraction of simulated experiments with maximum $-2\ln\lambda$ equal to or
greater than 14.4, and is 0.6\% which corresponds to the 2.5 $\sigma$ level of
excess over the background.
%%%%%%%%%%%%%%%%%%
%%%%%%%%%%%%%%%%%%

\begin{figure}[htbp]
\includegraphics[height=2.3in, width=3.5in]{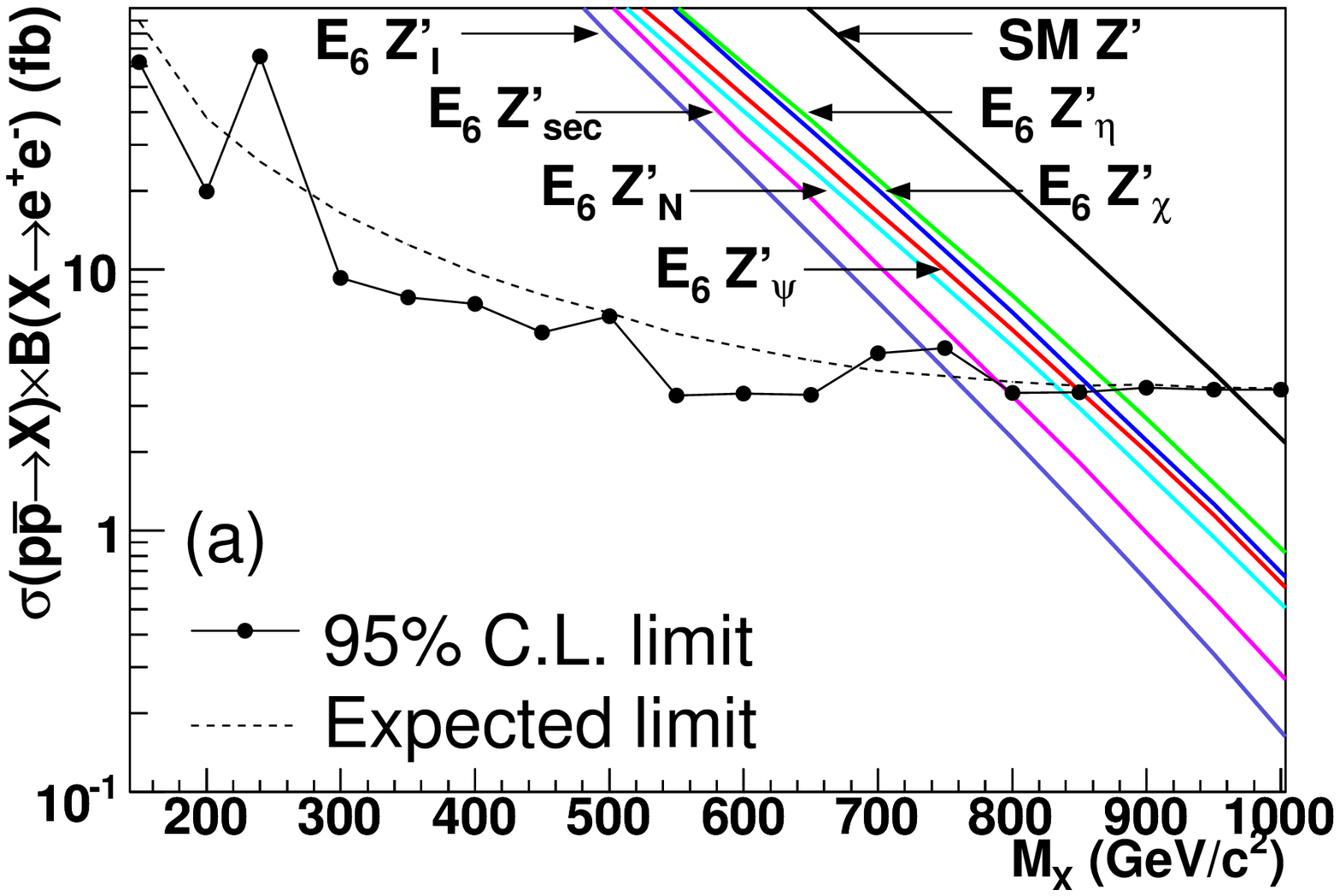}
\includegraphics[height=2.3in, width=3.5in]{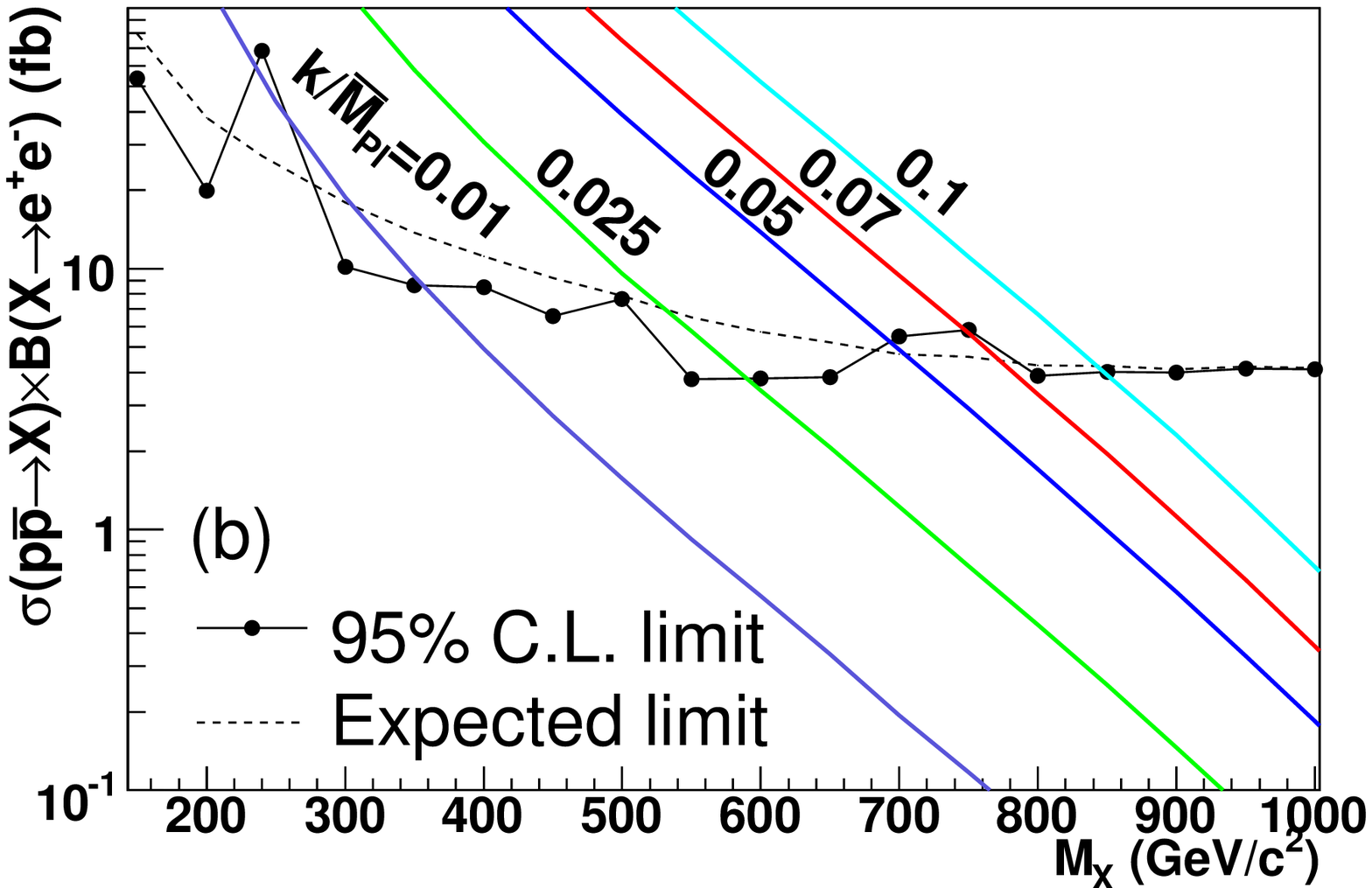}
\caption{The upper limits on $\sigma(p\bar{p}\rightarrow
  X)\cdot\mathcal{B}(X\rightarrow e^+e^-)$ as function of the mass of an $X$
  particle at the 95\% C.L. where $X$ is a spin 1 particle (a) or a spin 2
  particle (b) together with model predictions.}\label{fig:limits}
\end{figure}

\begin{figure}[htbp]
\includegraphics[height=2.3in, width=3.5in]{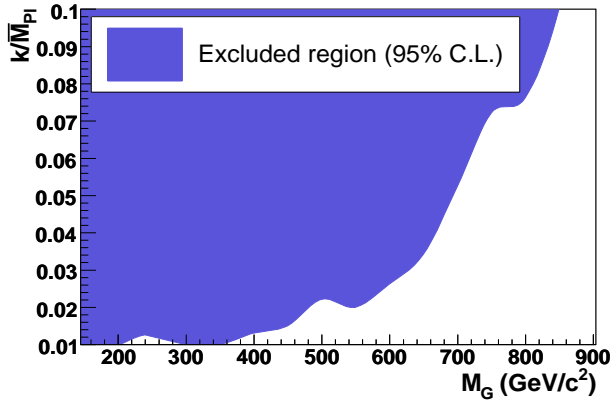}
\caption{$k/\overline{M}_{Pl}$ as a function of RS graviton mass. The shading
  indicates the region excluded at the 95\% credibility
  level.}\label{fig:rslimits}
\end{figure}

Upper limits on $\sigma(p\bar{p}\rightarrow X)\cdot\mathcal{B}(X\rightarrow
e^+e^-)$ at the 95\% C.L. are calculated as a function of mass using a Bayesian
binned likelihood method with a full consideration of uncertainties on the
total signal efficiency and the background
estimation~\cite{Joel}. Fig.~\ref{fig:limits} (a) shows the observed upper
limits from data and the expected limits from background-only simulated events
for spin 1 particles as a function of the $e^+e^-$ invariant mass, together
with the expected cross sections for
$Z'$s~\cite{zprime2}. Fig.~\ref{fig:limits} (b) shows the same but for spin 2
particles, together with the expected cross sections for RS gravitons. The
cross sections for $Z'$s and RS gravitons are calculated at leading order with
{\sc pythia} and then multiplied by a factor of 1.3 in order to approximate a
next-to-leading-order prediction as done in reports of earlier
results. Table~\ref{table:zprime} shows the lower mass limits of the SM
coupling and $E_6$ $Z'$s and Fig.~\ref{fig:rslimits} shows the excluded RS
graviton mass region with respect to $k/\overline{M}_{Pl}$.
%%%%%%%%%%%%%%%%%%
%%%%%%%%%%%%%%%%%%
\begin{table}[htbp]
  \begin{center}
  \caption{Expected and observed 95\% C.L. lower limits on $Z'$s masses.} \label{table:zprime}
  \begin{tabular}{cccccccc} \hline \hline
    $Z'$ Model &$Z'_{SM}$&$Z'_{\psi}$&$Z'_{\chi}$&$Z'_{\eta}$&$Z'_{I}$&$Z'_{sec}$&$Z'_{N}$ \\ \hline
    Expected Limit (GeV/$c^2$)&961&846&857&873&755&788&831 \\ 
    Observed Limit (GeV/$c^2$)&963&851&862&877&735&792&837 \\ \hline \hline
  \end{tabular}
  \end{center}
\end{table}
%%%%%%%%%%%%%%%%%%
%%%%%%%%%%%%%%%%%%

To conclude, we have searched for $e^+e^-$ resonances with 2.5 fb$^{-1}$ of
data collected by the CDF II detector. The largest excess over the standard
model prediction is at an $e^+e^-$ invariant mass of 240 GeV/$c^2$. The
probability of observing such an excess arising from fluctuation in the
standard model anywhere in the mass range of 150--1,000 GeV/$c^2$ is 0.6\%. We
also set upper limits on $\sigma(p\bar{p}\rightarrow
X)\cdot\mathcal{B}(X\rightarrow e^+e^-)$ at the 95\% C.L. for spin 1 and spin 2
particles. The SM coupling $Z'$ with mass below 963 GeV/$c^2$ and the $E_6$
$Z'$s with masses below 735/877 (lightest/heaviest) GeV/$c^2$ are excluded at
the 95\% C.L. RS gravitons with masses below 848 GeV/$c^2$ are excluded at the
95\% C.L. for $k/\overline{M}_{Pl}=0.1$.  

We thank the Fermilab staff and the technical staffs of the participating
institutions for their vital contributions. This work was supported by the
U.S. Department of Energy and National Science Foundation; the Italian Istituto
Nazionale di Fisica Nucleare; the Ministry of Education, Culture, Sports,
Science and Technology of Japan; the Natural Sciences and Engineering Research
Council of Canada; the National Science Council of the Republic of China; the
Swiss National Science Foundation; the A.P. Sloan Foundation; the
Bundesministerium f\"ur Bildung und Forschung, Germany; the Korean Science and
Engineering Foundation and the Korean Research Foundation; the Science and
Technology Facilities Council and the Royal Society, UK; the Institut National
de Physique Nucleaire et Physique des Particules/CNRS; the Russian Foundation
for Basic Research; the Ministerio de Ciencia e Innovaci\'{o}n, and Programa
Consolider-Ingenio 2010, Spain; the Slovak R\&D Agency; and the Academy of
Finland. 
%%%%%%%%%%%%%%%%%%

%%%%%%%%%%%%%%%%%%

\begin{thebibliography}{99}
\bibitem{cdfcoord}
  We use a cylindrical coordinate system where $\theta$ is the polar angle with
  respect to the proton beam axis (positive $z$ direction) and $\phi$ is the
  azimuthal angle. The pseudorapidity is $\eta=-\ln[\tan(\theta/2)]$. We define
  transverse energy as $E_T=E\sin\theta$ and transverse momentum as
  $p_T=p\sin\theta$, where $E$ is the energy measured in the calorimeter and
  $p$ is the magnitude of the momentum measured by the spectrometer.
\bibitem{ccong}
  Throughout this Letter the charge conjugate state is implied.
\bibitem{MET}
  Missing transverse energy (${\not\!\!E_T}=|{\not\!\!\vec{E}_T}|$) is defined as
  ${\not\!\!\vec{E}_T}=-\sum_i E^i_T \hat{n}_i$ where $\hat{n}_i$ is a unit
  vector in the transverse plane that points from the beam-line to the $i^{th}$
  calorimeter tower.
\bibitem{SOB}
  $\sigma_B$ is $\sqrt{N_B + \sigma^2_{N_B}}$, where $N_B$ is the number of
  expected backgrounds and $\sigma_{N_B}$ is the systematic uncertainty of the
  $N_B$.
\end{thebibliography}
\end{document}